\documentclass[prd,aps]{revtex4}
\usepackage{epsfig}

\newcommand{\dd}{{\rm d}}
\newcommand{\bk}{{\bf k}}
\newcommand{\bx}{{\bf x}}

\def\ii{{\rm i}}
\def\mG{{\cal G}}
\def\lg{\left\langle}
\def\rg{\right\rangle}
\def\lg{\left\langle}
\def\rg{\right\rangle}
\def\ds{{\delta\!s}}
\def\dsl{{\underline{\delta}\!s}}
\def\sigdsl{{\sigma_s}}

\begin{document}
\title{Non-Gaussianity in multi-field inflation}

\author{Francis Bernardeau \vskip 0.1cm}
\affiliation{Service de Physique Th\'{e}orique,
         CEA/DSM/SPhT, Unit\'{e} de recherche associ\'{e}e au CNRS, CEA/Saclay
         91191 Gif-sur-Yvette c\'{e}dex}
\author{Jean--Philippe Uzan \vskip 0.1cm}
\affiliation{Laboratoire de Physique Th\'{e}orique, CNRS--UMR 8627,
         B\^{a}t. 210, Universit\'{e} Paris XI, F--91405 Orsay Cedex,
         France,\\
         Institut d'Astrophysique de Paris, GReCO,
        CNRS-FRE 2435, 98 bis, Bd Arago, 75014 Paris, France.
\vskip 0.15cm}
\date{\today}

\begin{abstract}

This article investigates the generation of non-Gaussianity during
inflation. In the context of multi-field inflation, we detail a
mechanism that can create significant primordial non-Gaussianities in
the adiabatic mode while preserving the scale invariance of the power
spectrum. This mechanism is based on the generation of non-Gaussian
isocurvature fluctuations which are then transfered to the adiabatic
modes through a bend in the classical inflaton trajectory. Natural
realizations involve quartic self-interaction terms for which a full
computation can be performed. The expected statistical properties of
the resulting metric fluctuations are shown to be the superposition of
a Gaussian and a non-Gaussian contribution of the same variance. The
relative weight of these two contributions is related to the total
bending in field space. We explicit the non-Gaussian probability
distribution function which appears to be described by a single new
parameter. Only two new parameters therefore suffice in describing the
non-Gaussianity.
\end{abstract}
\pacs{{ \bf PACS numbers:} }
\vskip2pc

\maketitle
\section{Introduction}

The large scale structures of the universe are usually considered
to arise from vacuum quantum fluctuations that are amplified
during a stage of accelerated expansion. In its simplest version,
inflation predicts the existence of an adiabatic initial
fluctuation with Gaussian statistics and an almost scale-invariant
spectrum~\cite{inflation}. And it is clear that as long as the
evolution is linear from the radiation era, non-Gaussianity can
arise only from an ``initial'' non-Gaussianity generated during
inflation.

Simple calculations, that we reproduce in the second section of this
paper, however show that within single field inflationary framework it
is not possible to produce primordial non-Gaussian fluctuations if the
slow-roll conditions are preserved throughout the inflationary period
during which the seeds of the large-scale structures are
generated. Non-Gaussianity can be generated only if the inflation
starts from a non-vacuum initial state~\cite{martin} or if there exist
sharp features in the shape of the potential~\cite{staro}, but it in
the latter case it clearly shows up in the density fluctuation power
spectrum. It has been noticed however that the situation is somewhat
changed when more than one light scalar field are present during
inflation. In this case it is to be noticed first that one generically
produces a mixture of adiabatic and isocurvature
fluctuations~\cite{linde,polarski,garcia,ms} that can be uncorrelated
or correlated~\cite{langlois,gordon,hwang}. The observational
consequences of the existence of these two types of modes have started
to be considered~\cite{moodley1,lr,moodley2} but they clearly depend
on which type of matter each of the field decays to. Multi-field
inflation also opens the door to the generation of non-Gaussianity
simply because the non-linear couplings can be much stronger in the
isocurvature direction than in the adiabatic
direction~\cite{la,yamamoto,lm,salopek}. For instance in models in
which a Peccei-Quinn symmetry is broken during inflation, it was
pointed out by Allen {\em et al.}~\cite{allen} that a fourth order
derivative term in the effective theory leads to non-Gaussianity in
the axion density. In the "seed" models including
$\chi^2$~\cite{chi2}, axion~\cite{allen,axion}, Goldstone
bosons~\cite{bucher} and topological defects~\cite{topdef} models,
there is a test scalar field that is a pure perturbation and that does
not contribute to the background energy density; the energy being
quadratic in the field, it induces non-Gaussianity in the
perturbations. In such models however the non-Gaussian features are
present in the isocurvature modes only and will be observationally
relevant only if those modes survive the reheating phase. The
phenomenological situation is somewhat different however if a transfer
of the modes is possible, that is when the fields are
coupled~\cite{bartolo2,bartolo1}.

The aim of this paper is therefore to explore whether in the
context of multi field inflation it is possible to generate
non-Gaussian features while preserving the adiabatic slow-roll
type power spectrum and what would be its observational signature.
We are obviously motivated by the development of Cosmic Microwave
Background (CMB) and large scale structures observations that
offer an opportunity reconstruct the properties of the primordial
metric perturbations. Primordial fluctuations are more directly
probed by CMB observations but then the number of modes that can
be measured is still small. Up to now, no non-Gaussian signature
has been detected in either the 4 year COBE
data~\cite{sandvik,rocha} or the 1 year MAXIMA data~\cite{santos}.
In large-scale structure surveys the number of independent modes
one can observe is large but the difficulty is that the non-linear
gravitational dynamics~\cite{revue} generates non-Gaussian
couplings that can shadow the primordial ones~\cite{nous}.  One
should then rely on a good understanding of the impact of the
gravitational dynamics on the observations. Cosmic shear surveys
might offer one of the most serious opportunity to explore such
effects in the coming years~\cite{fb,detection}.

We start (Section~\ref{0}) by a general overview of the generation of
non-Gaussianity in single and multi-field inflation. It will lead us
to define a mechanism that can produce such non-Gaussianity. In
Section~\ref{I}, we then consider the evolution of perturbations in
two-field inflation and summarize how isocurvature perturbation can be
transferred to the adiabatic component when the trajectory in the
field space is curved. The isocurvature mode develops non-Gaussianity
due to self-interaction; we study in Section~\ref{II} the evolution of
such a self-interacting field in an expanding universe. After having
posed the problem for a quantum scalar field in an inflationary
background we address this issue from a classical point of view. In
Section~\ref{III} and ~\ref{PDF properties}, we compute the
probability distribution function that can be obtained in the class of
models considered in this article.

\section{Overview of the mechanism}\label{0}

As explained in the introduction, our goal is to design a model that
can produce sufficiently large non-Gaussianity at least for a band
$[\lambda,\lambda+\Delta\lambda]$ of wavelengths observationally
relevant, i.e. that corresponds to the large scale structure
scales. To guide us in this task, we review the properties of
non-Gaussianity in single field and multi-field inflation models in
order to determine whether they fulfill our requirements.\\

Let us start by considering a single field, $\phi$, in slow-roll
inflation. The Klein-Gordon equation for its perturbation is of
the form
\begin{equation}\label{ov:1}
\ddot{\delta\phi}+3H\dot{\delta\phi}-\frac{\Delta}{a^2}\delta\phi=-V''\delta\phi
-3V'''(\delta\phi)^2+\ldots
\end{equation}
During the slow-roll regime, $|\dot H|\ll H^2$ and
$\ddot\phi\ll3H\dot\phi$ so that
\begin{equation}\label{ov:2}
 3H^2M_4^2\simeq V,\qquad
 H\dot\phi\simeq -V',
\end{equation}
where $M_4$ is the reduced Planck mass, and the slow-roll conditions
can be expressed as $\varepsilon\ll1$ and $|\eta|\ll1$ with
\begin{equation}\label{ov:3}
  \varepsilon\equiv M_4\frac{V'}{V},\qquad
  \eta\equiv M_4^2\frac{V''}{V}.
\end{equation}
In order for the fluctuations of the scalar field to be large
enough, one needs the mass of the field to be much smaller than
$H$, in which case one gets $\delta\phi\sim H$. To
get the correct amplitude for the primordial fluctuations, one
needs $V^{3/2}/(V'M_4^3)\sim10^{-5}$ so that
\begin{equation}\label{ov:4}
 H\sim10^{-5}\varepsilon M_4.
\end{equation}
The band of wavelengths $[\lambda,\lambda+\Delta\lambda]$ corresponding to large scale
structures exits the Hubble radius during the $e$-folds
\begin{equation}\label{ov:5}
  N_\lambda=H\Delta t=\Delta\ln \lambda,
\end{equation}
during which the slow-roll parameter $\eta$ has varied as
$\Delta\eta\sim(\xi-\eta\varepsilon)\Delta\phi/M_4$, with
$\xi\equiv M_4^3(V'''/V)$. From Eq.~(\ref{ov:4}), one deduces that
\begin{equation}\label{ov:5bis}
 \Delta\phi/M_4\sim-\varepsilon H\Delta t.
\end{equation}
Now, if the quadratic term in the r.h.s. of Eq~(\ref{ov:1}) is
dominant over the linear term then $V''/V'''<\delta\phi\sim H$ and,
using Eq~(\ref{ov:4}), one deduces that $\Delta\eta>10^5\eta
N_\lambda$. This will induce a rapid breakdown of the slow-roll
inflation. This can be understood by the fact that the potential has
to be both flat enough for the fluctuations to develop and steep
enough for the non-linear terms not to be negligible. A way round to
this argument is to consider potential with a sharp
feature~\cite{staro} but in that case the non-Gaussianity is
associated with a departure from scale invariance and is located in a
very small band of wavelengths.\\

Thus, one requires at least one auxiliary field. As a second situation
let us consider the case in which this auxiliary field does not
interact with the inflaton so that the potential has the form
\begin{equation}\label{ov:6}
  V(\phi,\chi)=U(\phi)+\frac{1}{2}m^2\chi^2+\ldots
\end{equation}
so that the cosmological evolution drives $\chi$ towards the minimum
of its potential $\chi=0$. It follows that the lowest order
contribution to its energy density perturbation is not given by the
standard expression $\chi\delta\chi$ but one has to go to quadratic
order. Thus, even if the Klein-Gordon equation is linear,
the energy density of the field will develop non-Gaussianities with a
$\chi^2$ statistics and with $\rho_\chi\sim H^4$. This energy has to
be compared to the contribution of the inflaton $\rho_\phi\sim U\sim
H^2M_4^2$. It follows that $\delta\rho_\chi/
\delta\rho_\phi\sim10^{-5}(m/H)^{2}$, so that the contributions of the
auxiliary field to the background dynamics and to the Poisson equation
are negligible. $\chi$ can develop non-Gaussianities but they are not
transferred to the inflaton perturbation and gravitational
potential. \\

The two fields have to interact, a prototypal example being a
scattering term of the form $\mu\phi^2 \chi^2$ as proposed
in~\cite{lm}, so that there are a priori two sources of
non-Gaussianities: (i) as in the previous case, the density of the
field $\chi$ will be quadratic and (ii) due to the coupling, the
Klein-Gordon equation for the field $\phi$ will get an source term in
the r.h.s. of Eq.~(\ref{ov:1}). Consider a potential of the form
\begin{equation}\label{ov:7}
 V(\phi,\chi)=U(\phi)+\frac{1}{2}\mu\phi^2\chi^2+\frac{1}{2}m^2\chi^2
             +\frac{\lambda}{4!}\chi^4,
\end{equation}
so that the effective mass of the auxiliary field is $m^2_{\rm
eff}=m^2+\mu\phi^2$. If initially $|\phi|>|\chi|$, $\chi$ will roll
toward $\chi=0$ where it will stay so that $\phi$ drives the inflation
and $H^2M_4^2\sim U$. For the auxiliary field to develop non negligible
fluctuation, one needs that $m^2_{\rm eff}<H^2$. If $\mu>0$ then this
implies that both $m^2$ and $\mu\phi^2$ have to be smaller than $H^2$;
since during a period of $N_\lambda$ $e$-folds, using Eq.~(\ref{ov:5bis}),
the variation of the coupling contribution is of order $-2\mu\phi
M_4\varepsilon N_\lambda$, so that it can be smaller than $H^2$ for a number
of $e$-folds
\begin{equation}\label{ov:c}
 N_\lambda<10^{-5}H /(\mu\phi),
\end{equation}
during which $\delta\chi\sim H$. When $\mu<0$, the effective mass can
be smaller than $H$ due to cancellation between $m$ and $\mu\phi^2$
but still, the variation of the mass has to be smaller than $H$ so
that the condition (\ref{ov:c}) is a necessary condition whatever the
sign of $\mu$.

The Klein-Gordon equation (\ref{ov:1}) now has two source terms ${\cal
S}_1=\mu\chi\phi\delta\chi$ and ${\cal S}_2=\mu\chi^2\delta\phi$. The
solution will take the form $\delta\phi=\delta\phi_{_{\rm
H}}+\delta\phi_{_{\rm NG}}$ where $\delta\phi_{_{\rm H}}$ is the
solution of the homogeneous equation and $\delta\phi_{_{\rm NG}}$ a
particular solution. As will be detailed in Section~\ref{2b}, one
expects that at Hubble scale crossing $\delta\phi_{_{\rm NG}}\sim
N_\lambda{\cal S}/H^2$. It follows that for the first source term,
$\delta\phi_{_{\rm NG}}\sim N_\lambda\mu\phi\chi^2/H^2\sim
N_\lambda\mu\phi$. The amplitude of the homogeneous (Gaussian)
solution is $\delta\phi_{_{\rm H}}\sim H$ so that $\delta\phi_{_{\rm
NG}} \sim (\mu\phi/H) \delta\phi_{_{\rm H}} < 10^{-5}
\delta\phi_{_{\rm H}}$. The second source term gives rise to a
solution $\delta\phi_{_{\rm NG}}\sim
N_\lambda\mu\chi^2\delta\phi/H^2\sim N_\lambda\mu\delta\phi$; the
coefficient $\mu$ cannot be of order unity since otherwise it will
give a contribution of order $\mu\phi^2$ to the effective mass of
$\chi$. Thus if $\mu$ is of order unity either $m_{\rm eff}<H$ so that
$\chi\sim H$ but then it implies that $\phi/H<1$, which is not the
case during inflation or $m_{\rm eff}>H$ but then $\chi\ll H$. It
follows that in any situation the non-Gaussian correction is
negligible compared to the Gaussian contribution.

Now, let us study the relative magnitude of the different terms
entering the Poisson equation. First,
$U_{,\phi}\delta\phi\sim\varepsilon H^3M_4$ while the interaction term
is of order $V_{,\phi\chi} \chi\delta\phi\sim g^2\phi H^3 <
(10^{-10}/N_\lambda) \varepsilon H^3M_4$ and is thus
negligible. $V_{,\chi\chi}\chi^2$ has a contribution of order
$\lambda\chi^4\sim(\lambda 10^{-5}) \varepsilon H^3M_4$ which is
negligible and a second contribution $g^2\phi^2\chi^2< (10^{-10}/N_\lambda)
(\phi/H)\varepsilon H^3M_4$ which is also negligible. Note also that the
term $V_{,\chi}\chi\sim m^2_{\rm eff}\chi^2<H^4\sim10^{-5}\varepsilon
HM_4$, so that it is also subdominant.

It follows that even if $\chi$ develops non-Gaussianities and is
coupled to the inflaton, the requirement that its effective mass is
smaller than $H$ during a sufficient number of $e$-folds imply that it
is negligible both in the Poisson equation and in the inflaton
evolution equation.\\

As can be seen from the previous example, a quadratic interaction
does not fulfill the requirements. We turn to a toy model in
which the potential, in a neighborhood of the trajectory in field
space, takes the form
\begin{equation}\label{eq:toy}
  V=U(\phi)+m^2_\times(\phi-\phi_0)\chi + \frac{\lambda}{n!}\chi^n.
\end{equation}
Indeed, this very unusual form with linear coupling tends to show that
a particle physics realization of the scenario may be difficult to
build and the potential (\ref{eq:toy}) is just meant to be an
effective potential. As in the previous example, we have to estimate
the order of magnitude of the effect of $\chi$ both in the Poisson
equation and in the equation of evolution of the inflaton. Imposing
that the drift in the $\chi$ direction is smaller than in the $\phi$
direction during $N_\lambda$ $e$-folds gives that $N_\lambda<(H/m_\times)^2$. The
source term in the Klein-Gordon equation (\ref{ov:1}) of the inflaton
is $m^2_\times\chi$ so that it gives rise to a contribution
$\delta\phi_{_{\rm NG}}\sim N_\lambda m^2_\times/H<\delta\phi_{_{\rm H}}$
which can be of the same order as the homogeneous Gaussian solution
$\delta\phi_{_{\rm H}}$. In the Poisson equation, the only new
contribution is of order $V_{,\chi\phi} \delta\phi\chi \sim m^2_\times
H^2<(10^{-5}/N_\lambda)\varepsilon H^3 M_4$ and is thus negligible compared
to the contribution arising from $U$. As in the previous example, the
self-interaction term will also turn out to be negligible. As a
conclusion, such a potential will give rise to non-Gaussianities
sourced by the coupling to the inflaton perturbation. In such a
scenario, the field $\chi$ develops its non-Gaussianities due to
non-linear evolution and these non-Gaussianities are transferred to the
inflaton field due to the coupling source term in its Klein-Gordon
equation. Note that it does not generate directly non-Gaussianities in
the gravitational potential since its energy perturbation is always
negligible.\\

In a neighborhood of the minimum of the potential given by
$\chi_0(\phi)$, the potential can be developed in powers of
$\chi-\chi_0(\phi)\equiv\delta\chi$ as $\alpha_n\delta\chi^n$. The first
contribution is $\alpha_2\delta\chi^2$; either $\alpha_2>H^2$ and the
fluctuation of $\chi$ will be negligible or $\alpha_2<H^2$ and one
needs to consider next terms. If the cubic term is dominant then the
trajectory $\chi_0(\phi)$ is unstable. Now if it is either negligible
with the quartic order or of the same order, the latter situation
resulting in the fine tuning $\alpha_3\sim\alpha_4H$. For the
following $n>3$ terms, they will be associated with a particular
solution of magnitude $\delta\phi_{_{\rm
NG}}\sim\alpha_n\delta\chi^n/H^2\sim \alpha_nH^{n-4}\delta\phi_{_{\rm
H}}$. The $n$th order term will have a non negligible contribution to
$\delta\phi$ only if $\alpha_n\sim H^{4-n}$. Only for $n=4$ can
$\alpha_n$ be a pure number that can be chosen of order unity. This is
the only term for which the non-Gaussian fluctuations can develop on a
long time scale without any tuning to $H$. This naturality arguments
lead us to expect the most efficient and only relevant term to
consider is an auxiliary field self-interacting with a quartic
potential, $\lambda\chi^4/4!$, and coupled to the
inflaton~\footnote{Models of multi-field inflation~\cite{lyth} do not
generically present such properties. The construction of a reasonably
simple potential to realize this mechanism is left for future
investigation.}.

\begin{figure}
  \centerline{\psfig{figure=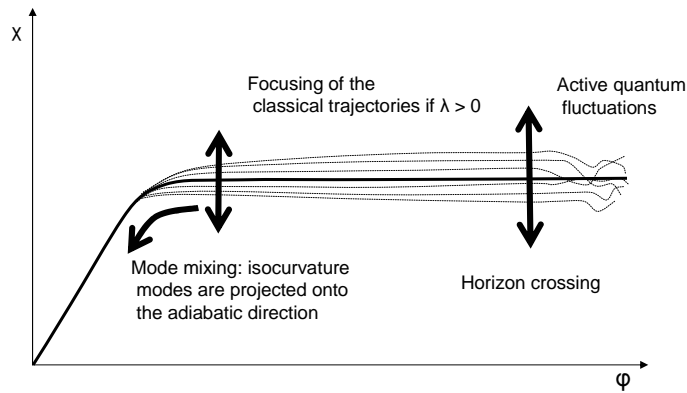,width=9cm}} \caption{The
  trajectories of the fields in the plane ($\phi,\chi$), once smoothed
  on a scale $R$. Before Hubble scale crossing ($R<H$), the
  trajectories behave quantumly because the quantum fluctuations are
  active up to scale $H$ and are not smoothed out. After the Hubble
  scale crossing ($R>H$), the trajectories can be treated as classical
  trajectories. Note that this transition happens at different time
  for different values of $R$. The bundle of classical trajectories
  then evolves in the two dimensional potential and its cross section
  evolves with time. During the bending of the potential valley, the
  isocurvature modes project on the isocurvature modes and there is a
  transfer because the length of each trajectory is different
  depending on its position.} \label{focus}
\end{figure}

The sketch of the scenario will thus be the following. The auxiliary
field develops non-Gaussianities. The bundle of field trajectories in
the field space $(\phi,\chi)$ is deformed when the potential minimum
is bent. According to the sign of $\lambda$, it is focalized
($\lambda>0$) or defocalized ($\lambda<0$) and the section of the
bundle is deformed, as depicted on figure~\ref{focus} (this completely
analogous to the deformation of the section of bundle of light
propagating in a gravitational field). Neighboring trajectories have
different lengths so that the fluctuations along each of the trajectories
will be slightly different.

To formalize this scenario, we describe in more details the coupling
and transfer between adiabatic and isocurvature modes. To characterize
the non-Gaussianity, one will need to treat the behavior of the
quantum fluctuation before Hubble scale crossing. But, the trajectories
after Hubble scale crossing can be treated classically and the classical
contribution is expected to dominate over the quantum one if the
bending occurs late enough after Hubble scale crossing.  Note that the case
in which $\lambda<0$ is less interesting since the trajectory becomes
unstable; $\chi_0(\phi)$ corresponds to a maximum of the potential and
we are more in situation like the one of a phase transition. We will
however consider the case $\lambda<0$ while computing the PDF.

\section{Two-field inflation: adiabatic-isocurvature transfer}
\label{I}

We describe, following mainly~\cite{gordon,hwang}, in this section
the mixing between adiabatic and isocurvature modes in multi-field
inflation which is one of the major element of our mechanism (see
also e.g.~\cite{kp} for an earlier study).

The class of multi-field inflationary models can be derived from a
Lagrangian for $N$ scalar fields
\begin{equation}\label{eq:lagrangien}
{\cal L}=-\frac{R}{16\pi
G}+\frac{1}{2}\sum_{j=1}^N\partial_\mu\varphi_j
\partial^\mu\varphi_j - V(\varphi_1,\ldots,\varphi_N)
\end{equation}
where $G$ is the 4-dimensional Newtonian constant, $R$ the Ricci
scalar and Greek indices run from 0 to 3. We describe the
universe by a Friedmann-Lema\^{\i}tre spacetime with metric
\begin{equation}\label{eq:metric}
\dd s^2=-\dd t^2+a^2(t)\delta_{ij}\dd x^i\dd x^j
\end{equation}
where $t$ is the cosmic time and $a$ the scale factor. We also
introduce the conformal time $\eta$ defined by $\dd t\equiv a\dd\eta$.
It follows that the Einstein equations take the form
\begin{eqnarray}
&&H^2=\frac{4\pi G}{3}\left(\sum_{j=1}^N\dot\varphi_j^2 +2V
\right)\label{eq:back1}\\
&&\dot H=-4\pi G\sum_{j=1}^N\dot\varphi_j^2\label{eq:back2}\\
&&\ddot\varphi_j+3H\dot\varphi_j=-V_j\label{eq:back3}
\end{eqnarray}
where a dot refers to a derivation with respect to the cosmic
time, $t$, $H\equiv\dot a/a$ and $V_{j}$ refers to a derivation
with respect to $\varphi_j$. We also set $M_4^{-2}\equiv8\pi G$.

Let us consider the simplest case in which we have only two scalar
fields $\varphi_1$ and $\varphi_2$. We decompose these two fields on
the direction tangent ($\sigma$)and perpendicular ($s$) to the
trajectory as
\begin{equation}\label{eq:rotation}
 \pmatrix{\dot \sigma \cr \dot s}
 =
 {\cal M}(\theta)
 \pmatrix{\dot\varphi_1 \cr \dot\varphi_2},\qquad
 {\cal M}(\theta)
 \equiv
   \pmatrix{
     \cos\theta & \sin\theta \cr
     -\sin\theta & \cos\theta}
\end{equation}
the angle $\theta$ being defined by
\begin{equation}\label{eq:def-theta}
\cos\theta\equiv\frac{\dot\varphi_1}{\sqrt{\dot\varphi_1^2+\dot\varphi_2^2}},
\qquad
\sin\theta\equiv\frac{\dot\varphi_2}{\sqrt{\dot\varphi_1^2+\dot\varphi_2^2}}
\end{equation}
from which it can be deduced that $s$ is constant along the
trajectory. The evolution of the field is obtained by
combining the two Klein-Gordon equations (\ref{eq:back3})
\begin{equation}\label{eq:evo-a}
  \pmatrix{\ddot\sigma \cr \ddot s}+ 3H \pmatrix{\dot \sigma \cr \dot s}
  + {\cal M}(\theta)\pmatrix{V_1 \cr V_2} = 0
\end{equation}
and the Friedmann equation is
\begin{equation}\label{eq:fried}
  H^2=\frac{4\pi G}{3}\left[\dot\sigma^2+2V(\sigma, s={\rm
  constant})\right].
\end{equation}

The interpretation of the fields $\sigma$ and $s$ as the adiabatic and
isocurvature components of $(\varphi_1,\varphi_2)$ will arise from the
following study of the perturbations evolution and properties. At
linear order, the perturbed metric takes the form~\cite{inflation}
\begin{equation}\label{eq:metric-pert}
\dd s^2=-(1+2A)\dd t^2+2a(t)\partial_iB\dd x^i\dd t
+a^2(t)[(1-2C)\delta_{ij}+2\partial_{ij}E]\dd x^i\dd x^j
\end{equation}
where $A$, $B$, $C$ and $E$ are four scalar perturbations and we
expand the scalar fields as
\begin{equation}\label{eq:pert-field}
  \varphi_i(t,x^j)=\varphi_i + \delta\varphi_i(t,x^j).
\end{equation}
We can introduce two sets of gauge invariant variables. The
Newtonian or longitudinal gauge is defined as
\begin{equation}\label{eq:newton}
  B=E=0,\qquad
  A=\Phi,\qquad
  C=\Psi,\qquad
  \delta\varphi_i=\chi_i.
\end{equation}
We introduce the perturbation of the inflaton in the flat slicing
gauge ($C=0, E=0$) by
\begin{equation}\label{eq:chgtjauge}
  Q_i=\chi_i+\frac{\dot\varphi_i}{H}\Phi.
\end{equation}
In Newtonian gauge, the evolution equations in Fourier space
reduce to the set
\begin{eqnarray}\label{eq:pert}
  &&\Phi =\Psi \\
&&\dot\Phi+H\Phi=\frac{1}{2M_4}(\dot\varphi_1\chi_1+\dot\varphi_2\chi_2)\\
  &&-\frac{\Delta}{a^2}\Phi+3H\dot\Phi +(3H^2+\dot H)\Phi=-\frac{1}{2M_4}
(V_1\chi_1+\dot\varphi_1\chi_1+V_2\chi_2+\dot\varphi_2\chi_2)\label{eq:poisson}\\
  &&\ddot\chi_i+3H\dot\chi_i-\frac{\Delta}{a^2}\chi_i+V_{ij}\chi_j=
  -2V_i\Phi+4\dot\varphi_i\Phi.\label{eq:kk}
\end{eqnarray}
By construction, $\ds$ is gauge invariant so that
$Q_s=\chi_s=\ds$ and from Eq.~(\ref{eq:chgtjauge}),
$Q_\sigma=\chi_\sigma+\dot\sigma\Phi/H$. Using that
\begin{equation}\label{eq:mat-prop}
  ^{t}\dot{\cal M}=-\dot\theta{}^{t}\dot{\cal M}J,\qquad
  ^{t}\ddot{\cal M}=-\ddot\theta{}^{t}\dot{\cal M}J
  -\dot\theta^2{}^{t}\dot{\cal M},\qquad
  J\equiv\pmatrix{0&1\cr -1&0},
\end{equation}
the Klein-Gordon equations (\ref{eq:kk}) can be rewritten as
\begin{eqnarray}\label{eq:kk2}
  &&\pmatrix{\chi_\sigma \cr \ds}^{..} +3H\pmatrix{\chi_\sigma \cr
\ds}^{.}
  +\left[-\frac{\Delta}{a^2}-\dot\theta^2+{\cal M}V_{ij} {}^{t}{\cal M}\right]
  \pmatrix{\chi_\sigma \cr \ds}=
  2\dot\theta J \pmatrix{\chi_\sigma \cr \ds}^{.}\nonumber\\
  &&\qquad\qquad\qquad\qquad\qquad
  +(\ddot\theta +3H\dot\theta) J\pmatrix{\chi_\sigma \cr \ds}
  -2\Phi{\cal M}\pmatrix{V_1 \cr V_2}+4\Phi\pmatrix{\dot \sigma \cr 0}
\end{eqnarray}
and the Poisson equation (\ref{eq:poisson}) takes the form
\begin{equation}\label{eq:poisson2}
 -\frac{\Delta}{4\pi
 Ga^2}\Phi=2\dot\theta\dot\sigma\ds+\ddot\sigma\chi_\sigma
  -\dot\sigma(\dot\chi_\sigma-\dot\sigma\Phi).
\end{equation}
Defining the mass matrix $U$ of the two fields $(\sigma,s)$ as
\begin{equation}
 \pmatrix{U_{\sigma\sigma} & U_{\sigma s} \cr
 U_{s\sigma} & U_{ss}} \equiv{\cal M}V_{ij} {}^{t}{\cal M},\qquad
 \hbox{and}\qquad
 \pmatrix{U_{\sigma} \cr U_{s}}={\cal M}\pmatrix{V_1 \cr V_2}
\end{equation}
and making use of (\ref{eq:poisson2}), the system (\ref{eq:kk2})
takes the form
\begin{eqnarray}
 &&\ddot\chi_\sigma+3H\dot\chi_\sigma+\left(-\frac{\Delta}{a^2}-\dot\theta^2+
 U_{\sigma\sigma}
 \right)\chi_\sigma =-2\Phi U_{\sigma}+4\dot\sigma\Phi
 +2(\dot\theta\ds)^.-2\frac{\dot\theta}{\dot\sigma}U_\sigma\ds\label{eq:systai}
\\
 &&\ddot{\ds}+3H\dot{\delta
s}+\left(-\frac{\Delta}{a^2}-\dot\theta^2+U_{ss}
 \right)\ds = -\frac{\dot\theta}{\dot\sigma}\frac{\Delta}{2\pi
 Ga^2}\Phi.\label{eq:systai2}
\end{eqnarray}
The comoving curvature perturbation is given by
\begin{equation}\label{eq:curv_pert}
  {\cal R}\equiv C+\frac{H}{\rho+P}(\dot\varphi_1\delta\varphi_1+
  \dot\varphi_2\delta\varphi_2)=C+\frac{H}{\dot\sigma}\delta\sigma
  =\frac{H}{\dot\sigma}Q_\sigma
\end{equation}
from which it is deduced that
\begin{equation}\label{eq:evoR}
  \dot{\cal R}=-\frac{H}{\dot
  H}\frac{\Delta}{a^2}\Phi+2\frac{H}{\dot\sigma}\dot\theta\ds.
\end{equation}

It follows from this analysis that as long as $\dot\theta=0$, $\ds$
evolves as a test scalar field evolving in an unperturbed
Friedmann-Lema\^{\i}tre universe and does not affect the evolution of
the gravitational perturbations; $\Phi$ couples solely to the
fluctuation of $\sigma$ in Eq.~(\ref{eq:poisson2}).  It is recovered
that the entropy remains constant on super-Hubble scales as long as
there is no mixing. $\ds$ will transfer energy to the gravitational
potential only when $\dot\theta\not=0$ and the comoving curvature can
be affected significantly even on super-Hubble scales if the fields
follow a curved trajectory. Note that this mechanism is effective only
during the inflationary period.

The ``surviving'' isocurvature perturbations (all of them in the case
where there is no bending) needs to be taken into account. At the end
of the inflationary period, both the inflaton and auxiliary field will
decay into particles and radiation.  To set the initial conditions in
the radiation era, one needs to specify in details these decays. When
there is no bending of the trajectory, this is the only mechanism
through which an imprint of the isocurvature modes can
survive. According to the scenario, the initial perturbations in the
cosmic fluids at the beginning of the radiation era are a mixture of
adiabatic and isocurvature modes that can be correlated (see
e.g.~\cite{langlois,gordon}). In full generality, we will have to
consider both contributions but we focus in the following on the
``pure'' case where the fluctuations at the end of inflation are
strictly adiabatic, which can occur if the two fields decay
identically.

\section{Generation of non-Gaussianity}
\label{II}

The goal of this section is to estimate the magnitude of the
non-Gaussianity developed by the isocurvature mode during a phase
of de Sitter inflation. The frame in which such calculations
should be performed is the quantum field theory for a coupled
scalar field. As we are interested in the emergence of weak
non-Gaussian features, through for instance the emergence of
non-zero connected part of high order correlation function, a
perturbation theory approach should be applicable in principle.

We point out however that a series of technical or conceptual
problems emerge in this physical situation for which there seems
to exist no known solution. We first present in which way the
computation we would like to do is affected by those problems but,
as their resolution goes far beyond the estimate we would like to
obtain, we finally turn to a classical treatment of the field
behavior. This provides a solid enough ground to estimate the
amounts of non-Gaussianities at scales that remain super-Hubble a
time long enough before the adiabatic-isocurvature mode transfer.

\subsection{The quantum level calculation}

As seen from the previous investigation, the entropy field $\ds$
is decoupled from the gravitational perturbations and can be
considered as a test field evolving in an homogeneous and
isotropic cosmological universe. To characterize the statistical
properties of such a self-interacting field, one would ideally
like to compute its high-order correlation functions. The
inflationary phase can be described by a de Sitter
spacetime~\cite{hawking} in flat spatial section slicing
\begin{equation}\label{eq:m}
\dd s^2=\frac{1}{(H\eta)^2}\left(-\dd\eta^2+\delta_{ij}\dd x^i\dd
x^j\right)
\end{equation}
which is conformal to half of the Minkowski spacetime. The conformal
time is related to the cosmological time by
\begin{equation}
\eta=-\frac{1}{H}\hbox{e}^{-Ht}
\end{equation}
and runs from $-\infty$ to 0, the limit $\eta\rightarrow0^-$
representing the ``infinite future''. The de Sitter spacetime can
be viewed as a four dimensional hyperboloid embedded in a five
dimensional Minkowski spacetime
\begin{equation}\label{eq:ds}
-\left(x_0\right)^2+\left(x_1\right)^2+\left(x_2\right)^2
+\left(x_3\right)^2+\left(x_4\right)^2=1/H^2.
\end{equation}
The invariance of this surface under five dimensional Lorentz
transformations implies that the de Sitter space enjoys a 10
parameter group of isometries known as the de Sitter group
$O(4,1)$. It is usual to define the de Sitter length function
$z(x,y)$ by $H^2(x-y)^2/2=1-z(x,y)$ where $x$ and $y$ are the five
dimensional coordinates of two points on the hyperboloid
(\ref{eq:ds}). It follows that $z(x,y)=H^2xy$ and the two points
are timelike or spacelike separated respectively when $z>1$ and
$z<1$

For a minimally coupled free quantum field of mass $m$, due to the
spatial translation invariance, the solution can be decomposed in
plane waves as
\begin{equation}\label{eq:quantdec}
 \widehat v_0(\bx,\eta)=\int\frac{\dd^3\bk}{(2\pi)^{3/2}}
 \left[v_0(k,\eta)\widehat b_\bk\hbox{e}^{i\bk\cdot\bx}
 +v^*_0(k,\eta)\widehat b_\bk^\dag\hbox{e}^{-i\bk\cdot\bx}\right]
\end{equation}
where we have introduced $\widehat v\equiv a\,\widehat\ds$, a hat
referring to an operator. In this Heisenberg picture, the field has
become a time-dependent operator expanded in terms of
time-independent creation and annihilation operators satisfying
the usual commutation relations $[\widehat b_\bk,\widehat
b_{\bk'}^\dag]=(2\pi)^3\delta^{(3)}(\bk-\bk')$. We can then define the
free vacuum state by the requirement
\begin{equation}
 \widehat b_{\bf k}\left|0\right>=0\qquad \hbox{for all}\qquad {\bf k}.
\end{equation}
As it is standard while working in curved space~\cite{birrel}, the
definition of the vacuum state suffers from some arbitrariness
since it depends on the choice of the set of modes $v_0(k,\eta)$.
They satisfy the evolution equation
\begin{equation}\label{eq:v}
  v_0''+\left(k^2-\frac{2}{\eta^2}-\frac{m^2/H^2}{\eta^2}\right)v_0=0,
\end{equation}
the general solution of which is given by
$\sqrt{\pi\eta/4}\left[c_1H^{(1)}_\nu(k\eta)+c_2H^{(2)}_\nu(k\eta)\right]$
with $|c_2|^2-|c_1|^2=1$, where $H^{(1)}_\nu$ and $H^{(2)}_\nu$
are the Hankel functions of first and second kind and with
$\nu^2=9/4-m^2/H^2$. Among this family of solutions, it is natural
to choose the one enjoying the de Sitter symmetry and the same
short distance behavior than in flat spacetime. This leads to
\begin{equation}\label{eq:sol}
v_0(k,\eta)= \frac{1}{2}\sqrt{\pi\eta}H^{(2)}_\nu(k\eta).
\end{equation}
This uniquely defines a de Sitter invariant vacuum state referred
to as the Bunch-Davies state vacuum~\cite{birrel}. In the massless
limit, the solution (\ref{eq:v}) reduces to
\begin{equation}
v_0(k,\eta)=\left(1-\frac{\ii}{k\eta}\right)\frac{\hbox{e}^{-\ii
k\eta}}{\sqrt{2k}}.
\end{equation}

Having determined the free field solutions, one can then aim to
express the $N$-point correlation functions of the interacting
field, $\ds_i$, in terms of those of the free scalar field. For
instance, on the example of a $\ds^4$ theory, the connected
part of the 4-point correlator of the interacting field will
reduce, at lowest order, to
\begin{equation}
 \left<0\right|T\widehat\ds_i(\bx_1,\eta_1)\ldots
 \widehat\ds_i(\bx_4,\eta_4)\left|0\right>=
 \ii\frac{\lambda}{4!}\int
 \left<0\right|T\widehat\ds(\bx_1,\eta_1)\widehat\ds(\bx_2,\eta_2)
 \widehat\ds(\bx_3,\eta_3)\widehat\ds(\bx_4,\eta_4)
 \widehat\ds^4(\bx,\eta)\left|0\right>\sqrt{-g}\,\dd^4\bx\,\dd\eta.
\end{equation}
Using the Wick theorem and keeping only the connected part leads
to
\begin{equation}
 \left<0\right|T\widehat\ds_i(\bx_1,\eta_1)\ldots
 \widehat\ds_i(\bx_4,\eta_4)\left|0\right>=
 \ii\frac{\lambda}{4!}\int \prod_{i=1}^4 G(\bx,\eta;\bx_i,\eta_i)
 \sqrt{-g}\dd^3\bx\dd\eta.
\end{equation}
$G(\bx,\eta;\bx',\eta')$ is the free propagator, that is the time
ordered product of two free fields in the free vacuum
\begin{equation}\label{eq:cor4}
\ii\,G(\bx,\eta;\bx',\eta')=\frac{1}{a(\eta)a(\eta')}\int\frac{\dd^3\bk}{(2\pi)^3}
\left[v_0(k,\eta)v_0^*(k,\eta')\hbox{e}^{-\ii\bk.\Delta\bx}\theta(-\Delta\eta)
+v_0(k,\eta')v_0^*(k,\eta)\hbox{e}^{\ii\bk.\Delta\bx}\theta(\Delta\eta)\right]
\end{equation}
with $\Delta\bx\equiv\bx'-\bx$ and $\Delta\eta\equiv\eta'-\eta$
and $\theta$ being the Heavyside function. After integration over
angles the free propagator can be computed~\cite{tsamis} to be
\begin{equation}\label{eq:propa}
\ii\,G(\bx,\eta;\bx',\eta')=\frac{H^2}{4\pi^2}\left(\frac{\eta\eta'}{\Delta\bx^2
  -\Delta\eta^2}+1+\int_0^\infty\frac{\dd k}{k}\cos(k\Delta x)
  \hbox{e}^{-\ii k\vert\Delta\eta\vert}\right).
\end{equation}
While trying to compute the 4-point correlator (\ref{eq:cor4})
with the latter expression of the free field propagator
(\ref{eq:propa}), one has to face the existence of two infrared
(IR) divergences. Note that at coinciding points, there is a UV
divergence that can be regularized by taking into account that
inflation started at a given initial time~\cite{phi2}.

The IR logarithmic divergence at $k=0$ was first exhibited by Ford
and Parker~\cite{irdiv}. Allen and Folacci~\cite{allenf} showed
that this divergence arises from the incorrect assumption of de
Sitter invariance for the vacuum and from the existence of a zero
mode, i.e. the action is invariant under transformations of the
form $\phi\rightarrow\phi+{\rm constant}$. It is well known that
an expansion in terms of creation and annihilation operators as in
Eq.~(\ref{eq:quantdec}) is inadequate for the zero
modes~\cite{dewitt} in the same way as the expansion in terms of
creation and annihilation operators for the standard quantum
harmonic oscillator breaks down when the frequency is going to
zero. To be slightly more precise, let us recall~\cite{fp} the
properties of a free massless scalar field leaving on a 3-torus of
volume $V$. Beside the standard plane wave solutions associated to
the creation and annihilation operators, $\widehat b_\bk$ and
$\widehat b_\bk^\dag$ obtained for $\bk\not=0$ and describing
harmonic oscillator of frequency $|\bk|$, a complete set of
solutions requires to consider the solution $(\widehat
x_0+\widehat p_0t)/\sqrt{V}$ obtained for $\bk=0$ and describing
the classical solution for a free particle. The position and
momentum operators $\widehat x_0$ and $\widehat p_0$ satisfy the
commutation relation $[\widehat x_0,\widehat p_0]=\ii$. A vacuum
state can then be defined by imposing $\widehat
p_0\left|0\right>=0$ and $\widehat b_\bk\left|0\right>=0$. This
state is the product of Fock and a Hilbert space corresponding
respectively to the oscillators and the free particle (see
e.g.~\cite{dewitt,fp,vilenkin,jaume,turok}) and will not be
normalizable. For a flat space in more than two dimensions, the
continuum limit exists because the contribution of the zero mode
is of zero measure, the volume of the phase space cancelling the
divergence but the effect remains in de Sitter space whatever its
dimension.

This divergence led some authors~\cite{allenf,jaume,polarski2} to
define other vacua with less symmetry than the de Sitter group but
with a well defined propagator. For instance, in the closed
spatial section slicing a natural choice is the $O(4)$ invariant
vacuum~\cite{allenf} that is symmetric under rotations of the
constant time hypersurfaces. The modes are discrete and the IR
divergence is avoided by choosing a set of modes with a different
solution for $\bk=0$. Kirsten and Garriga~\cite{jaume} proposed
the construction of an acceptable de Sitter invariant vacuum in
which the zero mode is well treated. The case of de Sitter space
with static spatial sections slicing was considered by
Polarski~\cite{polarski2}. In the case of flat spatial sections
slicing, which we are most interested in, the IR problem can be
regularized by working on a torus~\cite{tsamis} with
$-H^{-1}/2<x^i\leq H^{-1}/2$ which is equivalent to set an
infrared cut-off. With such a regularization, this yields the
result
\begin{equation}\label{eq:reg}
  \ii G(\bx,\eta;\bx',\eta')=\frac{H^2}{4\pi^2}\left(
  \frac{\eta\eta'}{\Delta\bx^2-\Delta\eta^2}
  -\frac{1}{2}\ln H^2(\Delta\bx^2-\Delta\eta^2)+ {\rm constant}\right).
\end{equation}

A second IR divergence arises at late time, i.e.  when
$\eta\rightarrow0^-$ since the integral (\ref{eq:cor4}) with the
regularized propagator (\ref{eq:reg}) diverges due to the volume
factor $1/H^4\eta^4$. The former resolution of the IR divergence
in $\bk=0$ cannot resolve this late time IR divergence. Tsamis and
Woodard~\cite{tsamis} pointed out that such correlation functions
suffers from a series of flaws: (1) they are not finite even at
lowest order, this problem becoming worth as the number of
vertices grows, (2) it is not purely imaginary so that it implies
a tree-order breakdown of unitarity. The physical origin seems to
be the redshifting that drives all physical momenta toward zero
when $\eta\rightarrow0$, making the overlap between plane waves
very strong. It is to be noted that the use of the regularized
propagator given in Eq. (\ref{eq:reg}) does not cure the problem
nor the introduction of a late time cut-off (the existence of
which could be associated with the reheating time).

The resolution of these fundamental problems goes far beyond the
estimates we want to obtain on the effects of nonlinear couplings.
We thus adopt a simpler approach assuming that at scales that
exceed the Hubble size the field value trajectories are classical
(e.g. deterministic) and encoded in the potential shape. Because
the trajectories might have a non-trivial dependence with the
initial field values set up at Hubble scale crossing, non-Gaussianities
can be induced during that period. If the time between Hubble
crossing and the bending of the trajectory is long enough the
non-Gaussianities of such classical origin are going to exceeds
those a priori present in the initial value distribution as tit
emerges from the quantum process.

\subsection{The classical limit}\label{2b}

To implement this idea, we consider the field $\ds$ at a large
enough scale. It amounts to applying to the evolution equation of the
field a filtering procedure at a fixed (comoving) scale, $R$. In
the following we note $\dsl$ the filtered field (what exactly is
the function it has been convolved with is not important). At the
time the Hubble size has shrunk below the smoothing length, the
trajectory of the field becomes classical. Its evolution equation
is simply,
\begin{equation}\label{dslevol}
  \ddot{\dsl}+3H\dot{\dsl}=\underline{\cal S}(\ds).
\end{equation}
It derives from the real space Klein-Gordon equation applied for
$\ds$ [e.g. Eq.~(\ref{eq:systai2})] where the Laplacian term has
been dropped on super-Hubble scales. Assuming that the trajectory
of the mean field value in a super-Hubble patch is insensitive to
the small scale fluctuations~\footnote{it amounts to neglect the
radiative corrections in the field evolution.} we can replace
$\underline{\cal S}(\ds)$ by ${\cal S}(\dsl)~$\footnote{In
Refs.~\cite{FP2}, the effect of the quantum subhorizon
fluctuations are described by a stochastic noise entering a
Fokker-Planck equation for the coarse-grained (or long wavelength
part) of $\ds$.}. In the absence of source terms in
Eq.~(\ref{dslevol}), $\dsl$ is simply constant (e.g. trajectories
are parallel lines on Fig.~\ref{focus}) and its value is given by
its initial value set up at Hubble crossing. The initial value is
given by a sum of Gaussian distributed values and non-Gaussian
corrections induced by the non-linear couplings during the
sub-Hubble evolution of the field,
\begin{equation}\label{dslzero}
  \dsl_{\rm init.}=\dsl^{(0)}+\dsl^{({\rm ng})}.
\end{equation}

The resolution of Eq. (\ref{dslevol}) requires in general the
knowledge of the source term. It can be solved however
perturbatively if one assumes that $\dsl$ can be expanded in terms
of the coupling constant entering the source term,
\begin{equation}\label{dslexpansion}
  \dsl(t)=\dsl^{(0)}+\dsl^{(1)}+\dots
\end{equation}
The evolution equation for $\dsl^{(1)}$ is trivial to obtain. It
reads,
\begin{equation}\label{dsl1evol}
  \ddot{\dsl}^{(1)}+3H\dot{\dsl}^{(1)}={\cal S}(\dsl^{(0)})
\end{equation}
the solution of which is
\begin{equation}\label{dsl1sol}
  \dsl^{(1)}(t)=\dsl^{({\rm ng})}+(t-t_{\rm H})\frac{{\cal
  S}(\dsl^{(0)})}{3\,H},
\end{equation}
which can be rewritten as
\begin{equation}\label{dsl1solNR}
  \dsl^{(1)}(t)=\dsl^{({\rm ng})}+N_R\,\frac{{\cal
  S}(\dsl^{(0)})}{3\,H^2},
\end{equation}
where $N_R$ is the number of $e$-folds from the Hubble size crossing, $t_H$,
to the time $t$. If the latter is large enough one expects the
second term of this equation to dominate the first, e.g. that
non-Gaussian effects induced during the classical evolution
dominates those built during the stage of the quantum evolution.
We will make this hypothesis in the following.

It clearly implies, as we anticipated in part~\ref{0}, that the
amplitude of the first order corrective term is given by the
amplitude of the source term ${\cal S}(\dsl^{(0)})$ divided by
$H^2$ times the number of $e$-fold during which the non linear
couplings are active.

The existence of such corrective terms are obviously of importance
for the statistical properties of the field. If the source term
contains nonlinear couplings the field $\dsl$ is no more Gaussian.
Such properties should be exhibited in the high order correlation
function of the field. Exploring these consequences is the aim of
the next section.

\section{Isocurvature modes Probability distribution function}
\label{III}

As seen from the previous analysis, nothing prevents the
isocurvature modes to develop non-Gaussian properties on
super-Hubble scales. As stressed before, if the time between
Hubble radius crossing and the exchange of modes is large enough, the
properties of those modes will be determined by their stochastic
evolution at super-Hubble scale, not so much by the quantum state
with which the field modes reach super-Hubble scales. In other
words we expect the high order correlation functions present in
the quantum field to be finally superseeded by the ones induced by
the subsequent stochastic couplings.

The aim of this section is then to characterize the way the
non-Gaussian features in the isocurvature modes are built up.
There are obviously many ways of characterizing non-Gaussian
features in a stochastic field. The simplest approach is to
consider the shape of the one-point probability distribution of
the local field value. This series depends obviously in the type
of couplings one has. For the reasons previously detailed,  we will
consider only the case of quartic couplings.

In the first part of this section (\S~\ref{sec:cumulant4}) we show
how it is possible to compute any of such cumulants at leading
order in the coupling constant (that would correspond to
tree-order calculation in a quantum field formulation). It finally
leads to the expression of the generating function of the
one-point cumulants~(\S~\ref{sec:cumulants}).

To get insights into the physical interpretation of the previous
results we then present (\S~\ref{subB}) the derivation of a
quantity more directly related to observations: the one-point PDF
of the local field value. Not surprisingly we will see that the
rare event tails of those distribution differs from those expected
for a Gaussian distribution. Depending on the sign of the quartic
coupling, one expects an excess or a deficit of rare values.

\subsection{The expression of the fourth order cumulant}
\label{sec:cumulant4}

Let us focus on the case of a {\it stochastic} field $\ds$
self-interacting with a potential,
\begin{equation}\label{vexp}
  V(\ds)={\lambda\over 4!}\,\ds^4
\end{equation}
and we assume that the coupling constant $\lambda$ is small
compared to unity. In the $\dsl$ expansion, Eq.~(\ref{dslexpansion}),
it means that $\ds^{(n)}$ is of the order of $\lambda^n$. We then
consider the evolution equation (\ref{dslevol}) with
\begin{equation}\label{quarticsource}
  {\cal S}(\dsl)=-\frac{\lambda}{3!}\dsl^3.
\end{equation}
At this stage it should be noted that when using the $\dsl$ field
in the source term, one neglects the effects of the small scale
fluctuations in the field trajectory. This is of no consequence
for tree order calculations, however, \emph{no loop terms, that
involve arbitrarily small scales fluctuations, can be reliably
computed from this approach}. If one wants to do that one should
solve the full quantum problem. In this study we allow ourselves
to use this simplification but forbid ourselves to compute
correlation properties beyond tree order.

As said before in the absence of coupling the free field solution
$\dsl^{(0)}$ is time independent. This is not the case of higher
order terms. For instance from Eq.~(\ref{dsl1solNR}), we know that
for this coupling term,
\begin{equation}
  \dsl^{(1)}=-{\lambda\over 18}\,N_R
  {\left[\dsl^{(0)}\right]^3\over H^2},
\end{equation}
a term which induces non-Gaussian properties in the field. In case
of a quartic coupling the value of $\dsl$ remains symmetric
distributed so that the third order cumulant is always zero. The
first non-vanishing high order cumulant is then the fourth one.
Its value can be computed at leading order (in the coupling
constant) from the expression of $\dsl^{(1)}$. Indeed the fourth
order cumulant is formally given by,
\begin{equation}\label{4ptcum}
\lg \dsl^4\rangle_c\equiv \langle \dsl^4\rg-3\lg \dsl^2\rg^2=\lg
\left[\dsl^{(0)}+\dsl^{(1)}+\dots\right]^4\rg-3\lg
\left[\dsl^{(0)}+\dsl^{(1)}+\dots\right]^2\rg^2,
\end{equation}
replacing $\dsl$ by its expansion in the coupling constant, Eq.
(\ref{dslexpansion}). The computation of these moments up to
linear order in $\lambda$ gives,
\begin{equation}\label{4ptcum2}
\lg \dsl^4\rg_c=\lg\left[\dsl^{(0)}\right]^4\rg-3
\lg\left[\dsl^{(0)}\right]^2\rg^2+4\left[
\lg\dsl^{(1)}\left[\dsl^{(0)}\right]^3\rg-3
\lg\dsl^{(1)}\dsl^{(0)}\rg \lg\left[\dsl^{(0)}\right]^2\rg
\right].
\end{equation}
This first two terms of this expression cancel because
$\dsl^{(0)}$ obeys a Gaussian statistics. One can finally check
that the latter expression gives the connected part of
$\lg\dsl^{(1)}\left[\dsl^{(0)}\right]^3\rg$ so that, at leading
order, the fourth order cumulant is given by,
\begin{equation}\label{4ptcum3}
  \lg \dsl^4\rg_c=4\lg\dsl^{(1)}\left[\dsl^{(0)}\right]^3\rg_c
\end{equation}
which can be easily computed from the expression of $\dsl^{(1)}$,
so that,
\begin{equation}\label{cumulant4}
  \lg \dsl^4\rg_c=-{4\lambda\over 3}\,{N_R\over H^2}
                           \lg\left[\dsl^{(0)}\right]^2\rg^3.
\end{equation}

\begin{figure}
  \centerline{ \psfig{figure=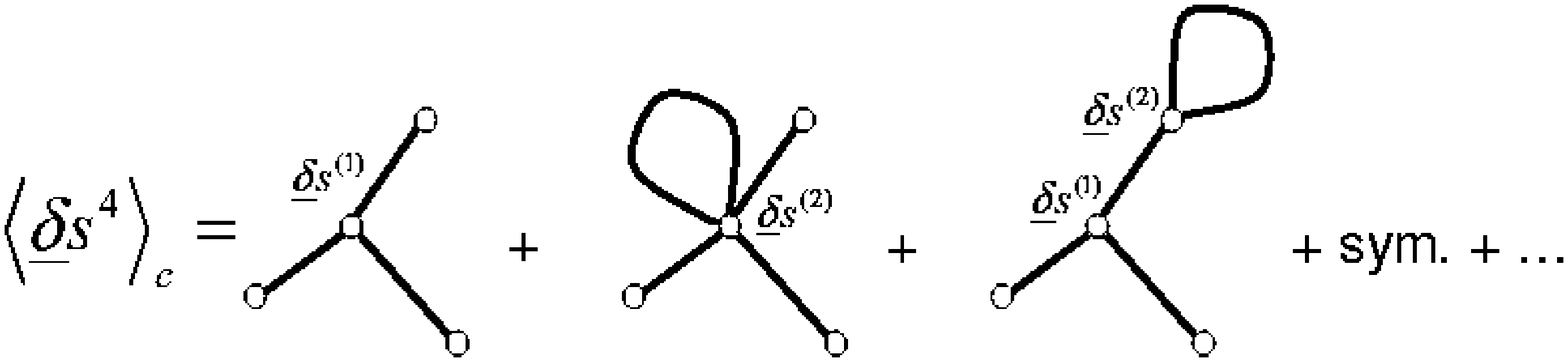,width=9cm}}
  \caption{Diagrammatic representation of the fourth order cumulant in a perturbation
  theory approach. Lines correspond to connected pair points in ensemble averages of
  product of Gaussian variables as an application of the Wick theorem. End
  points are taken at zero order in the coupling constant (they
  are linear in the initial Gaussian field). Points at first order
  in the coupling constant are cubic in the field, at second order
  they are quintic, etc.
  The first diagram corresponds to the tree order term. It involves
  only one vertex. The other two diagrams correspond to loop corrections.}
  \label{phi4c}
\end{figure}

It is clear that at leading order the expression of the fourth
order cumulant involves only the expression of $\dsl^{(1)}$. All
higher order terms in the expression of $\dsl$ contribute only at
higher order in $\lambda$ to this cumulant, and they would
correspond to loop term corrections, e.g. Fig. \ref{phi4c}.

\begin{figure}
  \centerline{ \psfig{figure=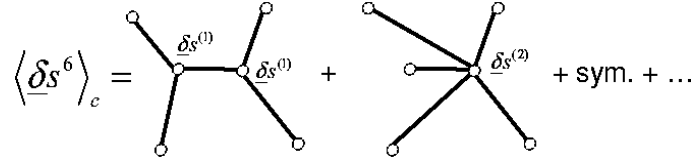,width=9cm}}
  \caption{Diagrammatic representation of the sixth order cumulant
  at leading order. Two types of diagrams appear, one with two
  three-leg vertices, one with one five-leg vertex.}
  \label{phi6c}
\end{figure}

However if one wishes to compute the expression of higher order
cumulants, such as the sixth order one, the higher order terms in
the expansion of $\dsl$ will then contribute (still at tree
order!), see Fig. \ref{phi6c}. In the next subsection we explore
in more details the mathematical structure of the cumulant
hierarchy.

\subsection{The structure of the field cumulants}
\label{sec:cumulants}

To get insights in how the leading order expression can be
computed, it is useful to consider one where terms of various
order in the expansion of $\dsl$ are mixed together.

For instance the sixth order cumulant is, at leading order in
$\lambda$, given  by
\begin{equation}\label{6ptcum}
\lg\dsl^6\rg_c=
30\,\lg\left[\dsl^{(1)}\right]^2\left[\dsl^{(0)}\right]^4\rg_c+
6\,\lg\dsl^{(2)}\left[\dsl^{(0)}\right]^5\rg_c
\end{equation}
since $\dsl^{(2)}$ scales like
$\lambda^2\left[\dsl^{(0)}\right]^5$. The factors that appear in
these expressions correspond to the number of each of such terms
that appear in those expansions.

In general the tree order term of any cumulant is determined by
the consequences of the Wick theorem applied to $\dsl^{(0)}$. It
will therefore be of the form,
\begin{equation}\label{cumulants}
  \lg \dsl^n\rg_c=\sum_{\rm decompositions}\lg\prod_{i=1}^n
  \dsl^{(p_i)}\dots\dsl^{(p_n)}\rg_c
\end{equation}
where all the decompositions are such that
\begin{equation}\label{decompositioncondition}
  \sum_{i=1}^n (2p_i+1)=2(n-1)
\end{equation}
so that it is possible to connect all terms together and no loop
can be built [$2(n-1)$ is two times the number of lines required
to connect $n$ points].

In order to manipulate dimensionless numbers we introduce the
(time dependent) vertices,
\begin{equation}\label{vertex}
  \nu_{2p+1}\equiv(2p+1)!\frac{\dsl^{(p)}}{\left[\dsl^{(0)}\right]^{2p+1}}.
\end{equation}
For instance
\begin{equation}\label{nu3val}
  \nu_3(t)=-\frac{\lambda\,t}{3\,H}=-\frac{N_R}{3\,H^2}.
\end{equation}
Then the high order cumulants can be rewritten as
\begin{equation}\label{cumulants2}
  \lg \dsl^n\rg_c=\sum_{\rm decompositions}\left[\prod_{i=1}^n
  \nu^{2p_i+1}\right]\ \lg \left[\dsl^{(0)}\right]^2\rg^{n-1}
\end{equation}
that appear to be identical to tree sums in which the $p$-leg
vertices are $\nu_p$.
The general computation of the $\nu_p$ is still a difficult task
since it requires the resolution of the evolution equation for
$\dsl$. Actually if the time after Hubble scale crossing is large
enough then the first term of the equation is finally negligible
so that the evolution equation actually reads,
\begin{equation}\label{mgevol2}
3H\dot\dsl(t)=-{\lambda\over 3!}\dsl(t)^3.
\end{equation}
Unlike the full evolution equation this equation has a simple
solution given by,
\begin{equation}\label{mgsol1}
  \left[\dsl^2(t)\right]^{-2}=\left[\dsl^{(0)}\right]^{-2}-{\lambda\,(t-t_{\rm H})\over 36\,H}
\end{equation}
the time dependence of which can be rewritten as a function
$\nu_3(t)$,
\begin{equation}\label{mgsol}
  \dsl(t)={\dsl^{(0)}\over\sqrt{1-\nu_3(t)\left[\dsl^{(0)}\right]^2/3}}.
\end{equation}
where $\dsl^{(0)}$ is the (time independent) value of $\dsl$ at
Hubble scale crossing. It is to be noted that the direct use of
this relation to compute the high order moments of $\dsl$ is
unjustified because it automatically introduces loop terms in the
computations. As these terms cannot be reliably computed we
restrict ourselves in the following to tree order terms. We can
also note that if $\nu_3$ is positive there is a singularity at
finite distance which would make all moment values infinite in
this case. This pathological behavior is due to the fact that on
rare occasions, the field value $\dsl$ moves at arbitrarily large
distance from the origin in a potential which is unbounded from
below. This instability is due to the description of the potential
we use around the trajectory and is not necessarily physical. Once
again if one limits ourself to tree order computations this issue
is automatically solved; all cumulants are finite are tree order.
This calculation rules provides us with a natural regularization
scheme.

A convenient way to describe what we have obtained is to write the
fourth order cumulant, in units of the second order moment,
\begin{equation}\label{4ptcumred}
  \frac{\lg \dsl^4\rg_c}{\lg \dsl^2\rg^3}=4\,\nu_3.
\end{equation}
The sixth order cumulant can similarly be expressed as a function
of $\nu_3$ and $\nu_5$
\begin{equation}\label{6ptcumred}
  \frac{\lg \dsl^6\rg_c}{\lg \dsl^2\rg^5}=(30\,\nu_3^2+6\,\nu_5).
\end{equation}
In general the cumulant of order $2n$ scales like $\lg \dsl^2\rg$
to the power $2n-1$ with a ratio given by a sum of product of
vertices.

It is useful to define the vertex generating function,
\begin{equation}\label{mgdef}
  \mG(t,\tau)\equiv\sum_p \nu_p(t)\frac{\tau^p}{p!},
\end{equation}
which can be straightforwardly related to the expression of $\dsl$
in Eq.~(\ref{mgsol}) when $\dsl^{(0)}$ is replaced by $\tau$ so that
\begin{equation}\label{mgsol2}
  \mG(t,\tau)={\tau\over\sqrt{1-\nu_3(t)\tau^2/3}}.
\end{equation}
This expression provides the values of all the vertices that can
be expressed in terms of $\nu_3(t)$, at least for large values of
$N_R$. We emphasize that this technique can be applied to any
potential shape, although it is probably not always possible to
find a close form for the generating function in all cases.

\subsection{The PDF calculation}\label{subB}

\subsubsection{General formalism}

The computation of the one-point PDF of the isocurvature
fluctuations relies on the use of its cumulant generating
function, $\varphi(y)$ defined below. All cumulants can obviously
be obtained, order by order, from the vertex generating function,
$\mG$, we have just obtained. But it is actually possible to take
advantage of their tree structure to compute the whole cumulant
generating function at once. This latter is defined as,
\begin{equation}\label{chidef}
  \varphi(y)\equiv\sum_n
  \frac{\langle \dsl^{2n}\rangle_c}{\lg \dsl^2\rg^{2n-1}} \frac{(-y)^{2n}}{2n!}
\end{equation}
This latter, because of the tree structure we are dealing with, is
obtained from a Legendre transform of the cumulant generating
function~\cite{revue},
\begin{equation}\label{varphi}
  \varphi(y)=y\,\mG(\tau(y))+{1\over 2}\tau^2(y)
\end{equation}
where $\tau(y)$ is solution of
\begin{equation}\label{taueq}
  \tau(y)=-y{\dd\mG(\tau)\over \dd\tau}.
\end{equation}
The derivation of the equation system is too long to be recalled
here (see~\cite{revue} for details). It is however worth noting that
\begin{equation}\label{chipexp}
  \frac{\dd\varphi(y)}{\dd y}=\mG(\tau(y)).
\end{equation}

\begin{figure}
  \centerline{ \psfig{figure=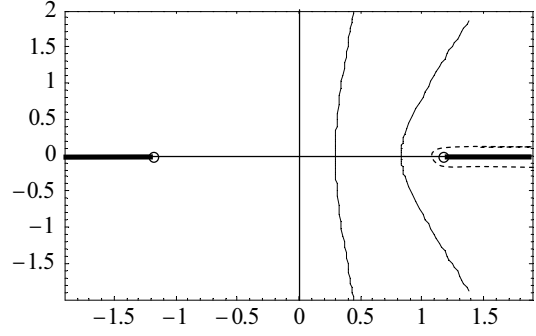,width=7cm}}
  \caption{Integration path for Eq. (\ref{pdfexp}) in the
  $y$-complex plane (thin solid lines) for $\nu_3\,H^2=0.3$ and $\dsl/\sigdsl=0.3$ and
  $1$. The two half bold lines on the real axis
  represent the location singularities for $\varphi(y)$. For large values of $\phi$ the integration
  path is pushed along the singular part (dashed line).}
  \label{ytraj}
\end{figure}

The one-point probability distribution function is then given by
the inverse Laplace transform of $\varphi(y)$
\begin{equation}\label{pdfexp}
  P(\dsl)=\int_{-\ii \infty}^{\ii \infty}\frac{\dd
  y}{2\pi\,\ii\,\sigdsl^2}\exp\left[ -\frac{\varphi(y)}{\sigdsl^2}
  +\frac{\dsl\ y}{\sigdsl^2}\right],
\end{equation}
where $\sigdsl$ is the variance of $\dsl$. The global properties
of $P(\dsl)$ can be derived from the properties of the cumulant
generating function.

{}From the expression (\ref{mgsol2}) one can obtain the expression
of $\tau$ in (\ref{taueq}),
\begin{equation}\label{taueq2}
  \tau(1-\nu_3\tau^2/3)^{3/2}=-y
\end{equation}
from which the expression of $\varphi(y)$ can be explicitly
computed. The properties of $\varphi(y)$ in the complex plane are
going to depend on those of $\tau$.

The shape of the PDF can be calculated from a numerical
integration in the complex plane. In practice to complete such a
numerical integration one must choose an adequate path in the $y$
plane to make the integral convergent. This is achieved in
imposing that the quantity in the exponent remains real along the
trajectory which can be obtained if the trajectory crosses the
real axis at the saddle point position, $y_s$, defined by
\begin{equation}\label{saddlepoint}
  \frac{\dd\varphi(y)}{\dd y}\vert_{y=y_s}=\dsl,
\end{equation}
which with Eq. (\ref{chipexp}) is given by the value of $\tau_s$
for which ${\cal G}(\tau_s)=\dsl$ and then
$y_s=-\tau_s/{\cal G}'(\tau_s)$. Example of such integration path are
presented on Fig. \ref{ytraj}.

\begin{figure}
\begin{center}
\psfig{file=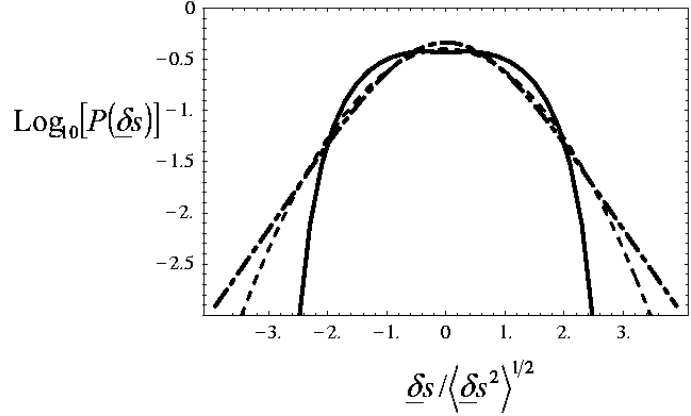,width=9cm}
 \caption{Shape of the one-point PDF of $\dsl$ for $\nu_3\,H^2=0.3$
 (dot-dashed line) or $\nu_3\,H^2=-0.3$ (solid line) compared to
 a Gaussian distribution (dashed line).}
 \label{ngplot}
\end{center}
\end{figure}

The resulting PDFs are shown on Fig. \ref{ngplot}. They clearly
exhibit non-Gaussian features (the Gaussian case is shown as a
dashed line). The magnitude of these features depend on the value
and sign of $\nu_3$, that is indeed related to the value and sign of the
coupling constant $\lambda$. In the following we explore in some
details the behavior of the PDF in different cases.

\section{Properties of the isocurvature mode PDF}
\label{PDF properties}

\subsection{The behavior of the PDF for small values of $\dsl$}

For small values of $\dsl/\sigdsl$ it is possible to derive an
explicit expression for the one-point PDF based on a saddle point
approximation. In Eq. (\ref{pdfexp}) one can expand $\varphi(y)$
around the saddle point defined previously, Eq.~(\ref{saddlepoint}).

The resulting formal expression of the PDF then reads,
\begin{equation}\label{pdfformulae}
  P(\dsl)\dd\dsl={\dd\dsl\over\sqrt{2\pi\varphi''(y_s)\,\sigdsl^2}}\,
  \exp\left[-{\tau^2\over 2\sigdsl^2}\right]
\end{equation}
which can be calculated from the explicit expression of $\varphi$,
\begin{equation}\label{pdfapprox}
  P(\dsl)\dd\dsl=
  \sqrt{\frac{3}{2\,\pi}\left|\frac{1-\dsl^2\nu_3}{(3+\dsl^2\,\nu_3)^3}\right|}
  \exp\left[-\frac{3\,\dsl^2}{(6+2\,\dsl^2\,\nu_3)\sigdsl^2}\right]
  \frac{\dd\dsl}{\sigdsl}.
\end{equation}
This result is valid as long as $\dsl^2\,\nu_3$ is small compared
to unity. For larger values of $\dsl$ the behavior of the PDF
depends crucially on the sign the $\lambda$.

This formula provides actually a very good description of the
overall PDF shape when $\lambda$ is positive (since excursion to
rare events are anyway not permitted).

\subsection{The rare event tails in case of negative $\lambda$}

When $\nu_3$ is positive, that is when $\lambda$ is negative, the
shape of the potential is such that it favors the rare event
tails. The behavior of the probability distribution function in
the rare event tails depends in particular on the analytic
properties of $\varphi(y)$ in the complex plane. For positive
values of $\nu_3$ it can be easily checked from Eq. (\ref{taueq2})
that $\tau(y)$, and consequently $\varphi(y)$ is non-analytic on
the real axis with two symmetric singularities at finite distances
of the origin,
\begin{equation}\label{tauc}
  \tau_c=\epsilon \,\frac{1}{2}\sqrt{\frac{3}{\nu_3}},
\end{equation}
($\epsilon=\pm 1$) corresponding to values of $\tau$ where its
derivative with respect to $y$ is diverging. In the vicinity of
this point it is easy to expand first the expression of $y$ as a
function of $\tau$,
\begin{equation}\label{yexpansion}
  y=\epsilon \,y_c+\epsilon \,\left(\tau-\tau_c\right)^2
\end{equation}
with
\begin{equation}\label{ycrit}
  y_c=-\frac{9}{16\sqrt{\nu_3}}
\end{equation}
and then the expression of $\varphi(y)$ reads [taking advantage of
Eq. \ref{chipexp}]
\begin{equation}\label{chiexpansion}
  \varphi(y)=\varphi_c+r_c\,(\epsilon \,y-y_c)-a_c(\epsilon\,
  y-y_c)^{3/2}+\dots
\end{equation}
with
\begin{eqnarray}\label{critvalues}
  \varphi_c&=&-\frac{3}{16\nu_3}\\
  a_c&=&\frac{1}{\sqrt{\nu_3}}\\
  r_c&=&\frac{16}{9\sqrt{3}\,\nu_3^{1/4}}
\end{eqnarray}
Those singularities induce exponential tails in the PDF of $\dsl$
(a very complete mathematical investigation of such cases can be
found in \cite{balian}). The shape of the tails are of
the form $\exp(-\vert\dsl\vert/\dsl_c/\sigdsl^2)$ where $\dsl_c$
is given by the inverse value of $y$ at the singularity,
$\dsl_c=16\sqrt{\nu_3}/9$.

To be more precise the expression (\ref{pdfapprox}) ceases to be
valid for $\vert\dsl\vert>\dsl_m$ where $\dsl_m\equiv
\mG(\tau_c)$. For larger values of $\dsl$ the saddle point
position in $y$ is pushed towards one of the singularities, $\pm
y_c$. The behavior of the PDF will then be dominated by the
behavior of $\varphi(y)$ around this point, Eq.
(\ref{chiexpansion}), in which a regular part
$\varphi_c+r_c(\epsilon\,y-y_c)$ and a singular part
$a_c\,(\epsilon\,y-y_c)^{3/2}$ appear. One can then make the
integration path in $y$ in Eq. (\ref{pdfexp}) running along the
real axis (both ways) crossing it at $y=y_c$ so that it can be
described by the real variable $u$ varying from 0 to $\infty$
with,
\begin{equation}
y=\epsilon\,y_c+\epsilon\,u\,e^{\pm \ii\pi}
\end{equation}
where the sign is changing according to whether $y$ is above or
under the real axis. Expanding the singular part in the
exponential one gets,
\begin{equation}
P(\dsl)= {a_c}\int_0^{\infty}\frac{\dd u}{2 \pi
\ii\,\sigdsl^4}\,u^{3/2} {e^{\pm 3\ii\pi/2}}
\exp\left[\frac{-\varphi_c+\epsilon\,y_c\,\dsl-(\epsilon\,\dsl-r_c)u}{\sigdsl^2}\right].
\end{equation}
The singular value in $y$ to consider (that is the value of
$\epsilon$ in the previous equation) depends on the sign of $\dsl$
(the positive tail corresponds to negative value of $y$) so that
finally one gets,
\begin{equation}
P(\dsl)={a_c\,\sigdsl\over \ \Gamma(-3/2)}
\left({\vert\dsl\vert-r_c}\right)^{-5/2}
\exp\left(-\frac{{\varphi_s}+\vert
y_s\,{\dsl}\vert}{\sigdsl^2}\right). \label{pdfcutoff}
\end{equation}
which, for the parameters describing the singular behavior of
$\varphi$, gives,
\begin{equation}\label{raretails}
  P(\dsl)=\frac{4\,\sigdsl}{3\sqrt{3\pi}\,\nu_3^{1/4}}\,
  \left(\vert\dsl\vert-\frac{1}{\sqrt{\nu_3}}\right)^{-5/2}
  \exp\left(\frac{3}{16\nu_3\,\sigdsl^2}-\left\vert\frac{9\dsl}
  {16\sqrt{\nu_3}\,\sigdsl^2}\right\vert\right)
\end{equation}
It is clear for this expression that the rare event tails are very
different from a Gaussian distributed variable.

\subsection{Bounding values in case of a positive $\lambda$}

When $\nu_3$ is negative, that is when $\lambda$ is positive, we
are in the opposite case. We expect that the rare event tails to
be chopped out. Actually in this case it is easy to see that there
is no singularity on the real axis. It implies that the
integration contour can be moved from the imaginary axis to the
left or the right to an arbitrarily large distance. When $\vert
y\vert\to \infty$, we have $\tau\sim \vert y\vert^{1/4}
(-3/\nu_3)^{3/8}$ which implies that,
\begin{equation}\label{philimit}
  \varphi(y)\sim -\left(\frac{3}{-\nu_3}\right)^{1/2}\vert y\vert
\end{equation}
for large values of $y$. As a consequence for large enough values
of $\phi$ the integral simply vanishes away. The values of $\phi$
are therefore are bounded by $\pm\sqrt{-3/\nu_3}$. The shape of
the PDF near the bounding values can also be computed explicitly
from the behavior of $\varphi$ for large values of $y$,
\begin{equation}\label{philimit2}
  \varphi(y)= -\left(\frac{3}{-\nu_3}\right)^{1/2}\vert
  y\vert+\frac{3^{3/4}}{(-\nu_3)^{3/4}}\sqrt{\vert y\vert}+\frac{9}{8\,\nu_3}\dots
\end{equation}
Then the expression of the integral (\ref{pdfexp}) reads,
\begin{equation}\label{pdfnearmax}
  P(\dsl)=\int_{-\ii\infty}^{+\ii\infty} \frac{\dd y}{2\pi\ii\,\sigdsl^2}
  \exp\left[\frac{(\sqrt{-3/\nu_3}-\vert \dsl\vert)\,\vert y\vert-a\,\vert y\vert^{1/2}-9/(8\nu_3)}
  {\sigdsl^2}\right]
\end{equation}
with
\begin{equation}
a=\frac{3^{3/4}}{(-\nu_3)^{3/4}}.
\end{equation}
A simple change of variable, $t^{1/2}=a\,y^{1/2}/\sigdsl^2$, shows
that it can be written,
\begin{equation}
P(\dsl)=\frac{\sigdsl^2}{a^2}\exp\left(-\frac{9}{8\nu_3\sigdsl^2}\right)
\int_{-\ii\infty}^{+\ii\infty} {\dd t\over
2\pi\ii}\exp\left(-t^{1/2}+z\,t\right)
\end{equation}
with
\begin{equation}
z={\sigdsl^2(\sqrt{-3/\nu_3}-\vert\dsl\vert)\over a^2}
\end{equation}

The behavior of the PDF near the bounding values are determined by
the small values of $z$. The expression of the PDF can be obtained
in this regime by a saddle point approximation, similar to Eq.
(\ref{pdfapprox}), which leads to,
\begin{equation}
P(\dsl)={\sigdsl^2\over a^2} {1\over2\sqrt{\pi}}\,
z^{-{3\over2}}\,\exp\left[-{1\over
4\,z}-\frac{9}{8\nu_3\sigdsl^2}\right].
\end{equation}
which reexpressed in terms of $\dsl$ gives the behavior of the PDF
near the bounds,
\begin{equation}
P(\dsl)={1\over2\sqrt{\pi}\,\sigdsl}\,
\left(1-\frac{\vert\dsl\vert}{\sqrt{-3/\nu_3}}\right)^{-{3\over2}}\,\exp\left[-{9\over
8\nu_3\,\sigdsl^2}-{(-3/\nu_3)^{3/2}\over
4\,\sigdsl^2(\sqrt{-3/\nu_3}-\vert\dsl\vert)}\right].
\end{equation}
The PDF is found to go continuously to zero at $\dsl=\pm
\sqrt{-3/\nu_3}$ positions.

\section{Conclusions}

In this paper we have explored the possibility of generating
significant non-Gaussian initial metric perturbations in the
context of inflationary cosmology while preserving a power
spectrum of slow roll type adiabatic fluctuations. We found that
the only viable mechanism is through a multiple field inflation
where transverse (e.g. isocurvature) modes developed non Gaussian
properties that can be subsequently transferred to the adiabatic
fluctuations if the classical field trajectory is bent.

We have pointed out that quartic type coupling in the transverse modes is the most
natural type of couplings for providing non-Gaussianities in the sense
that in this case no fine-tuning in the value of the coupling constant
has to be invoked. We stress that in the context of such a mechanism,
unlike any others, the amount of non-Gaussianities that can be fuelled
in the adiabatic fluctuations can be almost arbitrarily large.

We have examined in more details the case where the isocurvature
mode generation (that is when they reach the Hubble size during the
inflationary period) and the adiabatic-isocurvature mode mixing
happen at very different time. The reason we consider this case is
two-fold. First it is somewhat pedagogical since it shows that
these two stages do not have to be concomitant. Second it implies
that the non-Gaussian properties of the isocurvature modes
developed mainly during the time they live at super-Hubble
scales. It makes their computation much more simple since we can
avoid a full treatment of the nonlinear field evolution at a
quantum level (and it turns out that such a computation is not
straightforward at all!). For modes living at super-Hubble scales
we allowed ourselves to view the field evolution as the one of a
classical stochastic field. In this case it is then possible to
pursue the calculations to completion in the sense that it is
possible to derive their whole set of correlation properties.

In particular we have been able to derive the one-point field
cumulants at tree order in the weak coupling limit in a consistent way
and finally build up the one-point probability distribution function
of the field value. In such class of models, the statistical
properties of the curvature perturbation are described by the
superposition of a Gaussian and a non-Gaussian contributions with a
relative weight proportional to the bending angle, $\Delta\theta$, of
the trajectory in field space during slow-roll. The non-Gaussian
component is fully characterized by a single parameter, $\nu_3$,
related to the reduced fourth order connected cumulant and has the
same variance as the Gaussian contribution.  For practical purposes,
we emphasize that its PDF is well approximated by
Eq.~(\ref{pdfapprox}) that reproduces accurately the numerically
computed PDF within our approximation scheme. Thus, all the
statistical properties can be encapsuled in two parameters.  These
results give some insights on what type of non-Gaussian features can
appear in future large-scale structure or CMB surveys while assuming
the inflationary prejudice. Note that the kind of non-Gaussianity
described here departs from that generated by the non-linear
gravitational dynamics, in particular it has no skewness.  There is
still however some ways between these results and their observational
consequences. How non-Gaussian properties that are present at
super-Hubble scales are transferred for instance to the CMB
anisotropies at sub-Hubble scales is not totally
straightforward. Whether such effects could be actually observed when
observational aspects are taken into account demands in-depth
analysis.

\section*{Acknowledgements}

We thank Robert Brandenberger, Nathalie Deruelle, Lev Kofman, Ian
Kogan, Jean Iliopoulos, Jihad Mourad, Renaud Parentani, Simon
Prunet, Alain Riazuelo and Richard Schaeffer for discussions and
the Institut d'Astrophysique de Paris for hospitality.


\end{document}